\documentclass[aps,twocolumn,prb]{revtex4}

\usepackage{graphicx}
\usepackage{dcolumn}

\begin{document}

\title{Enhancement factor distribution around a single SERS Hot-spot and
its relation to Single Molecule detection}

\author{E. C. Le Ru} \email{Eric.LeRu@vuw.ac.nz}

\author{P. G. Etchegoin} \email{Pablo.Etchegoin@vuw.ac.nz}

\author{M. Meyer}

\affiliation{The MacDiarmid Institute for Advanced Materials and
Nanotechnology, School of Chemical and Physical Sciences, Victoria
University of Wellington, PO Box 600, Wellington, New Zealand}

\date{\today}

\begin{abstract}
We provide the theoretical framework to understand the phenomenology
and statistics of single-molecule (SM) signals arising in
Surface-Enhanced Raman Scattering (SERS) under the presence of
so-called electromagnetic hot-spots (HS's). We show that most
characteristics of the SM-SERS phenomenon can be tracked down to the
presence of tail-like (power law) distribution of enhancements and
we propose a specific model for it. We analyze, in the light of
this, the phenomenology of SM-SERS and show how the different
experimental manifestations of the effect reported in the literature
can be analyzed and understood under a unified ``universal''
framework with a minimum set of parameters.
\end{abstract}


\maketitle

\section{Introduction}
\label{SecIntro}

The possible detection of a Single Molecules (SM) by Surface
Enhanced Raman Scattering (SERS) has played an important role in
reviving the interest in SERS. \cite{1997Nie,1997Kneipp} However,
most of the evidence originally put forward in favor of SM-SERS were
indirect. For each piece of evidence, one could usually find an
alternative explanation that did not require SM-SERS to be
real.\cite{1999Chimia} Despite these problems, the consensus is that
SM-SERS is indeed possible. Although each piece of evidence in favor
of SM-SERS is not a proof in itself, the accumulated body of
evidence from different experiments gives, as a whole, a strong
argument that it must be real. There remains the question however of
how to determine, for a given SERS experiment, whether we are in a
regime of single molecule SERS or not. This is a very important
problem, since the interpretation of many recent SERS experiments
rely on the fact that SM-SERS is observed, but do not provide an
unambiguous proof that it is indeed the
case.\cite{2006ShegaiJPCB,1999MichaelsJACS} One pre-requisite to
observe SM-SERS is the existence of positions of large SERS
enhancements, so-called hot-spots (HS's) on a SERS substrate.
Suitable HS's are believed to be formed for example at a junction
between two metallic nanoparticles,
\cite{2000MichaelsJPCB,2000XuPRE,2006FutamataFD} and are highly
localized. \cite{2004CPLHS} This HS localization makes it very
difficult to design experiments where SM-SERS signals are observed
and unambiguously demonstrated.

The ideal approach would be to take one molecule and place it
precisely at a HS, but there is currently no easy way of achieving
this. Laser forces on the probe molecule could be an option but it
has not yet been demonstrated experimentally to
occur.\cite{2006KallFD} The probe molecule(s) therefore adsorb at a
random position on the substrate. The most common approach currently
relies on an ultra-low concentration of analyte, which ensures that
one molecule (at most) will adsorb at the HS and in principle
guarantees the single molecule nature of the dectected signals. This
approach presents several shortcomings. Firstly, it heavily relies
on a correct estimate of the surface coverage of the analyte. This
is often prone to errors, and can be particularly deceitful for low
concentration where wall adsorption during preparation or
contamination can lead to large errors in the analyte concentration.
Secondly, the probability that the single molecule adsorbs at a HS
(the only place where it can be detected) is very small. This leads
to very unreliable statistics (since most HS's are unoccupied) that
would not pass even the simplest tests of statistical confidence. In
practice, this approach may (at best) provide an indication of the
presence of HS's with single molecule detection capabilities, but is
not suited to carry out systematic SERS experiments using single
molecules. In our opinion, the most convincing method based on
ultra-low concentrations is possibly the one based on
Langmuir-Blodgett monolayers, developed by Aroca and
co-workers.\cite{2001ConstantinoAnChem} While suffering still from
some of the drawbacks regarding the statistics of observed
intensities, it is one of the methods that offers the largest degree
of control over where the dyes are and how they are distributed over
the sample.

Another possible approach is to use larger analyte concentrations to
ensure that most HS's are occupied by one molecule at least (on
average). This approach also has its drawbacks. Firstly, we need to
address the question of how we know the concentration that is
required to have one molecule at HS's on average. Secondly, even if
we know the answer to the latter question, we now have many more
molecules at non-HS positions on the substrate. A natural question
then arises: How do we know if the SERS signal is dominated by these
many molecules with a low EF, or by the one at the HS with a large
EF? In many cases, the presence of fluctuations or of a Poisson
distribution of intensities are taken as evidence for SM signals,
and these questions are then ignored. In fact, these two questions
are strongly intertwined with the EF distribution. By studying this
distribution and its implication on the statistics of SERS signals,
we will show that these arguments (Intensity fluctuations and
Poisson distributions of intensity) are not always correct and need
to be assessed carefully. An alternative method, which we recently
developed, relies on the simultaneous use of two distinguishable
analytes to answer these questions, at least
experimentally.\cite{2006JPCBBiASERS} This method is simple to
implement and provides a general recipe for proving the presence of
SM-SERS signals under different experimental conditions. Moreover,
this technique showed that the SERS signals from small colloidal
clusters in liquids can be dominated by a few molecules only, even
at fairly high concentration. This conclusion highlights the
importance of the distribution of SERS enhancements in SM-SERS
experiments and it is one of the aims of this paper to study this
connection.

The distribution of enhancements in metallic structures is relevant
to many other techniques exploiting local field enhancements, such
as Surface Enhanced Fluorescence (SEF) and, to some extent, to
plasmonics in general. Previous studies have concentrated mainly on
the maximum local field enhancements that can be achieved, in
particular at hot-spots. This approach is important, for example, to
understand whether enhancements are sufficiently large to observe
signals from single molecules. It is appropriate if one can easily
place a molecule {\it exactly at the HS}. In most cases, however,
molecules are randomly positioned, and it is therefore equally
important to study what happens {\it around the HS} and with the
overall distribution of enhancements.

We emphasize in this work the connection between the SM-SERS problem
and the distribution of SERS enhancement factors (EF's) on a given
substrate. Large SERS enhancements ($\approx 10^8-10^{10}$) are
required to detect a single molecule. There are currently no known
SERS substrates that exhibit such large enhancements uniformly
across the substrate. On the contrary, these large EFs are believed
to occur precisely at HS's, typically located in a narrow gap
between two metallic
objects.\cite{2000MichaelsJPCB,2000XuPRE,2006FutamataFD} This can
for example be at a junction between two metallic particles, or
between a metallic tip and a metallic substrate in Tip-Enhanced
Raman Scattering (TERS) experiments. Within this picture, the HS
covers a small surface area compared to the rest of the substrate,
but exhibits a much higher EF and should therefore contribute
substantially to the signal. We will here quantify more carefully
this assertion and put it in the context of several claims made in
the past in the field of SM-SERS.

The paper is organized as follows: in Sec. \ref{SecEF}, we study the
SERS enhancement distribution around a single hot-spot and discuss
its main characteristics. The discussion is based on a
representative example, and we propose a simple analytical model for
the enhancement distribution. In Sec. \ref{SecHStoHS}, we discuss
how the EF distribution may change depending on the HS
characteristics. In all cases, the main property of the EF
distribution around a HS is its ``long-tail'' nature. In Sec.
\ref{SecStats}, we discuss the implications of these ``long-tail''
distributions for the statistics of SERS signals, in particular in
the single molecule regime.

\section{Enhancement Factor distribution at a gap hot-spot}
\label{SecEF}

\subsection{Presentation of the problem}

In order to study how the enhancement distribution may affect the
statistics of SERS signals, we first need to find the distribution
itself in a situation where HS's are present and, in particular, in
the case of one single HS. The strongest HS's, and therefore the
ones required for SM detection, are believed to be ``gap HS's'',
formed at a junction between two closely spaced metallic objects.
\cite{2006FutamataFD} We will therefore study a model system of two
closely spaced spherical metallic particles.
\cite{2006PCCPAbs,2006PCCPPol} This system presents a strong HS in
between the two particles, as required, and captures essentially the
physics of most ``gap'' HS's, formed by two closely spaced objects.
The conclusions will therefore remain valid for more complex
systems, qualitatively and
semi-quantitatively.\cite{2006PCCPAbs,2006PCCPPol}

This electromagnetic problem can be solved analytically in 3D
following Generalized Mie Theory as developed in Ref.
\onlinecite{1982GMT}, and the electromagnetic enhancements can
therefore be calculated pseudo-analytically to a high accuracy.
The series and matrices required for its solution were computed
numerically, and we checked the convergence of the solution. The
calculated EF's are therefore exact. This is important since many
of the approximations used to solve such problems tend to have
problems when estimating the local field enhancements, especially
at HS's. This is because most numerical techniques, such as Finite
Element Modelling\cite{FEM1,FEM2,FEM3} or the Discrete Dipole
Approximation,\cite{Purcell,Yang,FFT,Markel} rely on a
discretization (meshing) of the geometry. This gives satisfactory
results in terms of far-field properties (scattering or
extinction), but rapidly face limitations when computing the local
field at the surface. This is because of the large field gradients
existing at a HS, which require the use of an extremely fine mesh,
often resulting in prohibitively large computational time for 3D
simulations. 2D simulations can be useful in understanding
qualitative effects \cite{2Dsimulations}, but are unlikely to
predict realistic enhancement factors for 3D structures.

The SERS electromagnetic enhancement factors, $F(\mathbf{r})$ can be
calculated at any position from the knowledge of the local field
$\mathbf{E}_{\mathrm{Loc}}(\mathbf{r})$ (the incident field
amplitude is denoted $E_0$). It depends in principle on the Raman
tensor of the probe, its adsorption geometry, the scattering
geometry (e.g. backscattering), and the energy of the vibrational
mode.\cite{2006LeRuCPLE4} To avoid unnecessary complications, we use
the approximation $F \approx |E_{\mathrm{Loc}}/E_0|^4$, which is
simpler, more general, and sufficient for our purpose here. The
meaning of this approximation has been discussed in Ref.
\onlinecite{2006LeRuCPLE4}. We will focus primarily on $F$, which
represent the electromagnetic SERS enhancement factor, but it is
easy to adapt our results to $M=\sqrt{F}$, i.e. local field
intensity enhancement, which is important in other situations, such
as enhanced absorption or SEF.

\begin{figure}
\centering{
\includegraphics[width=8cm]{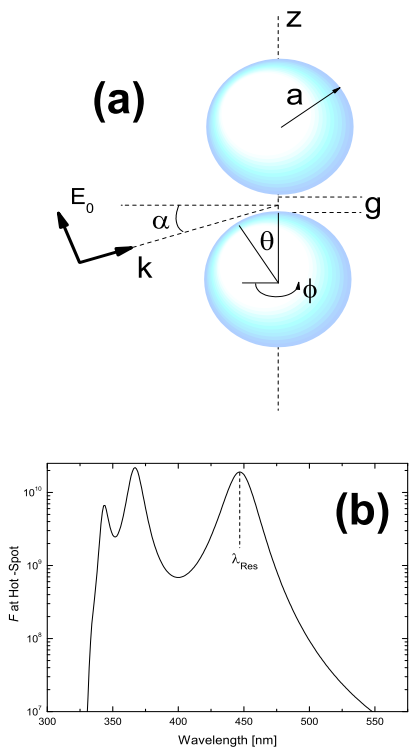}
} \caption{(a) Schematic illustration of the electromagnetic problem
under consideration: a dimer formed by two silver particles of
radius $a=25$\,nm and separated by a gap $g=2$\,nm, excited by a
plane wave polarized along the dimer axis $(z)$, i.e. $\alpha=0$.
Any position on the surface of the bottom particle are defined by
standard spherical coordinates angles $\theta$ (measured from the
dimer axis) and $\phi$ (measured from the incidence plane). (b)
Calculated enhancement factor $F=|E_{\mathrm{Loc}}/E_0|^4$ at the
surface of the bottom particle along the dimer axis (i.e. at the hot
spot, $\theta=0$). The resonance at $\lambda_{\mathrm{Res}}=448$\,nm
is a plasmon resonance resulting from the dipolar interaction
between the two particles. The other peaks are due to higher order
interaction. Note that $F$ remain large over a wide range of
wavelengths.}
 \label{FigLambdaDep}
\end{figure}

To illustrate our discussion, we will consider the specific case of
a dimer of two silver spheres in air excited in a configuration as
shown schematically in Fig. \ref{FigLambdaDep}(a). The polarization
of the incident field is aligned along the dimer axis ($\alpha=0$).
This is the configuration where the largest enhancement is predicted
at the HS in the gap. The parameters used are as follows: radius
$a=25$\,nm, gap $d=2$\,nm, and the local dielectric function of
silver was used.\cite{Claro} We show in Fig. \ref{FigLambdaDep}(b)
the wavelength dependence of the EF at the HS, i.e. at the surface
of one of the particles along the dimer axis in the gap. The
resonance due to dipolar interaction between the two spheres is the
most red-shifted one, and occurs at $\lambda\approx 448$\,nm. It
would be further red-shifted to $\approx 542$\,nm in water and even
further if the particles were not spheres. This resonance is the
important one for SERS, and we therefore choose the excitation to be
at $\lambda=448$\,nm to study the EF distribution. The results are
largely independent of this choice, as we shall see later.

Finally, we will only focus in this work on the distribution of EF's
on the metallic surface. We therefore study the EF experienced by
molecules directly adsorbed on the surface and ignore any effects of
additional layers of molecule. It is generally believed that this
first layer contributes to most of the SERS signal\cite{Otto}.
Moreover, in situations of single molecule detection, the
concentration is well below monolayer coverage, hence justifying
even further the choice. Note that we also neglect any chemical
contribution to the SERS enhancement.

\subsection{Enhancement distribution at the surface}

\begin{figure}
\centering{
\includegraphics[width=8cm]{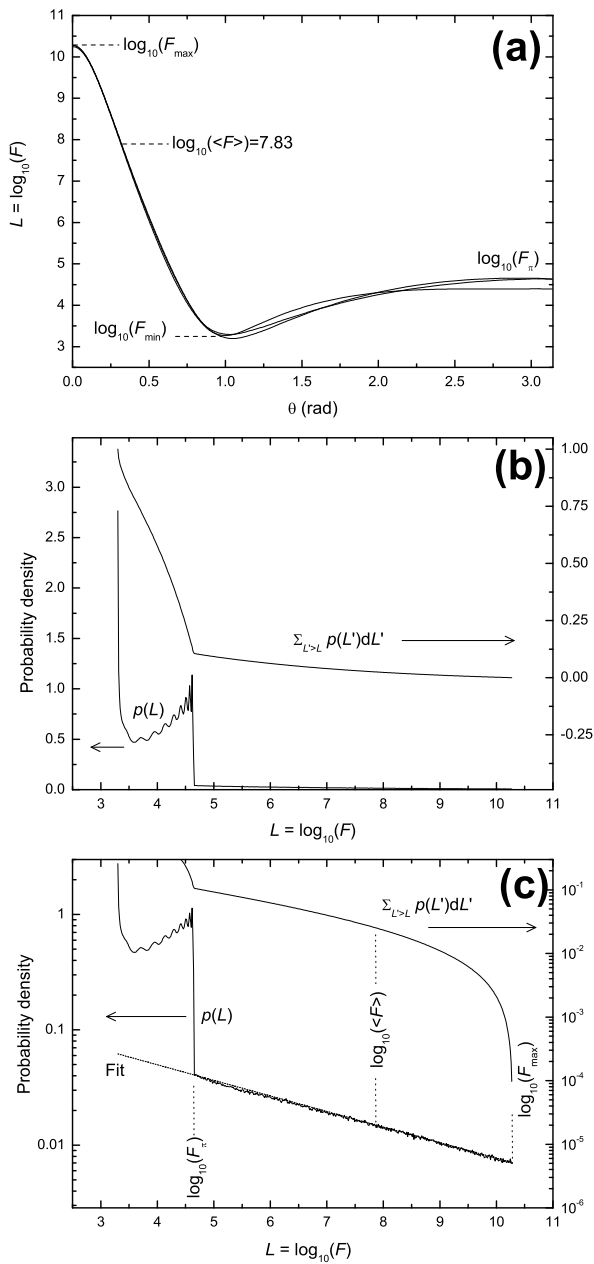}
} \caption{Several representations of the EF distribution on the
surface of one sphere of a dimer. Due to the many orders of
magnitude spanned by $F$, the EF is characterized by $L=\log_{10}
F$. (a) $L$ as a function of $\theta$ for $\phi=0$ and $\phi=\pi/2$.
Also shown is the case $\alpha=45^\circ$ where the incident wave
comes at a 45$^\circ$ angle with polarization at 45$^\circ$ from
dimer axis (EF scaled here by a factor 4, see Sec. \ref{SecHSPol}).
The three curves are nearly identical, especially close to the HS
($\theta=0$). (b) Probability distribution function (pdf) of $L$,
$p(L)$, for a random distribution of molecules on the sphere. $p(L)$
can be derived from $L(\theta)$ taking into account the fact that
$p(\theta)=(1/2)\sin\theta$. The quantity $p(L'>L)=\sum_{L'>L}
p(L')dL'$ (i.e. 1-cdf , where cdf is the cumulative distribution
function)  is also shown. The distribution is so skewed that the
interesting region of large enhancement ($L>5$) can hardly be seen.
(c) Same as (b) on a semi-log scale to highlight the tail of large
enhancements. The fit to the pdf of a truncated Pareto distribution
is also shown (dotted line).}
 \label{FigDistr}
\end{figure}

We calculated $F$ for positions on the surface defined by $0\le
\theta \le \pi$ (see Fig. \ref{FigLambdaDep}(a)) and various $\phi$
(note that the symmetry of revolution is broken here by the
direction of the wave-vector of the incident beam, so there could be
a $\phi$ dependence). Because of the large variation of $F$, it is
more convenient to characterize the EFs by $L=\log_{10} F$. We show
the distribution $L(\theta)$ in Fig. \ref{FigDistr}(a) for $\phi=0$
and $\phi=\pi/2$. It is clear that the results are virtually
identical, especially in the region of interest (close to the HS).
We will therefore ignore the $\phi$ dependence in the following:
$L(\theta,\phi) \approx L(\theta,\phi=0)$. We also see in
\ref{FigDistr}(a) a very sharp increase in the EF as one approaches
the HS ($\theta=0$). The EF drops by a factor of 10 from $\theta=0$
to $\theta=0.19$\, rad ($11^\circ$). The maximum $F$ in this example
is $F_{\mathrm{max}}=1.9\times 10^{10}$ at the HS, while the minimum
is $1.5\times 10^3$ for $\theta=\pi/3$. This clearly highlights the
huge variation in $F$ across the substrate, over 7 orders of
magnitude here.

From the curve $L(\theta)$, we can in principle derive the EF
distribution on the surface, i.e. the probability density function
(pdf) $p(F)$ that a molecule at a random position experiences a
given enhancement $F$ (strictly speaking $p(F)dF$ is the probability
that the enhancement is between $F$ and $F+dF$). To do so, we need
to take into account the fact that the problem is in 3D with
(approximate) symmetry of revolution. A given $\theta$ corresponds
to a ring on the sphere surface, whose radius varies with $\theta$.
Therefore, if we pick a random position on the surface, $\theta$ is
not uniformly distributed, but has a pdf $p(\theta)=(1/2)
\sin\theta$. The pdf $p(F)$ for $F$ can be expressed analytically as
a function of $F(\theta)$. However, because we do not have here an
analytical expression for $F$, but only its value for discrete
$\theta$'s, it is easier to determine $p(F)$ using a
Monte-Carlo-type approach. We generate random $\theta$ according to
their probability distribution $p(\theta)$ and for each of them
calculate the corresponding $F$ using a quadratic interpolation of
the curve $F(\theta)$ around $\theta$. $p(F)$ and $p(L)$ are then
obtained by drawing a (normalized) histogram of the obtained $F$ and
$L$. Various representations of this distribution are shown in Fig.
\ref{FigDistr}(b-c). Note that we have to generate a large number of
random $\theta$ ($10^7$ here) to produce smooth plots because the
probability distribution is very skewed and spans a wide range. The
average enhancement per molecule can be obtained in two equivalent
ways, either using $F(\theta)$ or $p(F)$:
\begin{equation}
\langle F\rangle=\int_0^\pi F(\theta)\frac{\sin\theta}{2}d\theta =
\int F p(F)dF.
\end{equation}
Both lead to $\langle F\rangle=6.7\times 10^7$ in this case
($\log_{10}\langle F\rangle=7.83$). The average of the distribution
is explicitly shown in Fig. \ref{FigDistr}(a) and it shows that it
is completely dominated by the region close to the HS itself. The EF
distribution shown in Fig. \ref{FigDistr}(a-c) captures all the
important aspects of SERS enhancement around a single HS and their
relation to statistical fluctuations, as we shall show in what
follows.

\subsection{Principal characteristics of the EF distribution at a HS}

 We will now discuss several general
important points. The main characteristic of the distribution $p(F)$
is that it is a so-called {\it ``long-tail'' distribution}. It is
similar in some ways to the Pareto distribution, encountered in many
areas of physics, economy, and social sciences.\cite{Pareto} It is
often used as a paradigm to model the wealth distribution in a
society.\cite{Pareto2} Fig. \ref{FigDistr}(a-c) actually reveal that
SERS enhancements are some sort of an extreme example of a Pareto
distribution (compared with the examples found in social
sciences)\cite{Pareto2}, because the enhancement is completely
dominated by what happens in a very narrow angular range. We
summarize the main characteristics of this distribution here:

\begin{itemize}
\item
 The average SERS enhancement is $\langle F\rangle=6.7\times 10^{7}$, which
is about 285 times smaller than the maximum enhancement
$F_{\mathrm{max}}=1.9\times 10^{10}$. This means that a single
molecule exactly at the HS would contribute as much to the SERS
signal as $\sim 300$ randomly positioned molecules.
\item
 For a
uniform distribution of molecules on the surface, 80\% of the SERS
signal originates from only 0.64\% of the molecules. These
correspond to the molecules in a small disk-like region around the
HS characterized by an angular spread of $\theta < 0.16$\,rad
($9^\circ$). An equivalent version of the Pareto
principle\cite{Pareto2} for our system would be that: on average,
98\% of the SERS signals originates from only 2\% of the (randomly)
adsorbed molecules.
 \item
 These considerations have immediate
implications in terms of fluctuations of the signal.
 If one records
the SERS signal in successive events, each originating from say 300
randomly adsorbed molecules, then strong fluctuations are
automatically expected. This is because on average the signal will
be dominated by only 0.64\% of these, i.e. 2 molecules. Fluctuations
are therefore bound to arise depending on where exactly in the HS
these two contributing molecules are adsorbed. We shall return to
this issue later.
 \item
 The distribution has such inequalities that the contribution from molecules
 far from the HS is always negligible. Despite the fact that there
 could be a large number of them (say 1000 times more than close to
 the HS), their enhancement factor is so small (at least $10^5$
 smaller) that their contribution can be safely ignored. In other
 words, the exact form of the distribution $p(L)$ for say $L < 6$ has
 no impact on most observations.
\end{itemize}

\subsection{A simple model hot-spot}

This last remark is in fact quite useful, because it enables us to
model the distribution with analytical expressions with only 3
parameters. First, we see in Fig. \ref{FigDistr}(c) that the tail
for large $L$ can be well-approximated by a straight line. This
means that:
\begin{equation}
\log_{10} p(L) \approx -k L +c.
\end{equation}
Using the fact that $p(F)=p(L)/(F\ln(10))$, this leads to the
empirical formula:
\begin{equation}
p(F) \approx A F^{-(1+k)}.
\end{equation}
In our example, we extract $k\approx 0.135$ and $A\approx 0.075$.
The corresponding fit is shown in Fig. \ref{FigDistr}(c). This
expression is similar to that of the pdf for a Pareto distribution.
There is a difference, however, in the value of $A$ because our
distribution does not extend to $F\rightarrow\infty$, but has a
maximum value: $p(F)=0$ for $F>F_{\mathrm{max}}$, corresponding to
the maximum physically achievable enhancement in the sample.
Therefore, it corresponds to a {\it truncated Pareto distribution}
(TPD). Under these conditions, we can approximate the enhancement
distribution by a TPD with 3 parameters: $k$, $A$, and
$F_{\mathrm{max}}$. This distribution is very accurate for large
enhancements $L>5$, but not for lower enhancements (see Fig.
\ref{FigDistr}(c)). We have already emphasized, nevertheless, the
fact that lower enhancements have a negligible contribution, so this
approximate distribution is in fact excellent for most predictions,
regardless of how the distribution is extended to lower
enhancements.

The approach we describe is, in fact, a very general approach for
characterizing SERS enhancements in a substrate containing HS's. All
important physical predictions can be deduced from the value of
these three parameters. The mathematical details of such derivation
are presented in Appendix \ref{AppA}. In our example, we have
$k=0.135$, $A=0.075$, and $F_{\mathrm{max}}=1.9\times 10^{10}$. With
these parameters, we can (for example) re-derive from Eq.
(\ref{EqnAveF1}) the previously obtained average of $\langle F
\rangle = 6.7\times 10^7$. We now discuss briefly the physical
meaning of these three parameters for a given HS:
\begin{itemize}
\item
 $F_{\mathrm{max}}$ simply describes the maximum intensity of the
 HS and can be viewed as the ``strength'' of the hot spot.
 It is the enhancement experienced by a molecule placed exactly
 at the right spot. This needs to be sufficiently large if
 SM-signals are to be detected.
\item
 $A$ is an indication of how probable it is for a molecule to be located
 at (or close to) the HS. In a way, $A$ is not really a characteristics of the HS
 itself but more of the remaining metallic surface. It is a
 representation of the relative area of the HS with respect to the
 rest of the substrate.
 In simple terms, the larger the value of $A$, the larger the
 hot-spot area with respect to the rest. The current description
 (with a truncated Pareto distribution) would fail though if $A$
 is too large, since the HS would then not have enough
 ``contrast'' in enhancement with respect to the rest in order to
 be considered as a proper hot-spot.
 \item
 $k$ determines how fast the enhancement decreases when moving away
 from the HS. It is therefore (indirectly) a measure of the sharpness of the
 resonance (in spatial terms) of the HS. To visualize this, let us consider for
 example the number of molecules experiencing an enhancement
 $F=F_{\mathrm{max}}/10$. These molecules are all located at the same angle
 $\theta$ from the HS (which is $\theta=0$ in Fig. \ref{FigLambdaDep}(a)), i.e. they form a ring around the HS.
 The number of these molecules increases with $\theta$. It should also be proportional to
 $p(F_{\mathrm{max}}/10)=A(10/F_{\mathrm{max}})^{1+k}$, which is decreasing with $k$.
 When $k$ increases, the corresponding ring must then be smaller (i.e. smaller $\theta$'s), and
 the spatial resonance is sharper.
 A larger $k$, for the same $A$ and $F_{\mathrm{max}}$, therefore corresponds to
 a sharper resonance, with larger enhancement gradients.
Note however, that for general values of $k$, $A$ and
$F_{\mathrm{max}}$, the sharpness of the resonance is better
characterized by the value of $D$ given in Eq. (\ref{EqnAveF2})
(large $D$ for sharper resonances).
\end{itemize}

\subsection{Enhancement Localization}

It is often stated in the SERS literature that HS's are places of
highly localized enhancements on the surface. We can now quantify
this assertion using our model HS.
 We define the $q$-HS ($0\le q \le 1$) as the region around the
 HS from which a proportion $q$ of the total SERS signal originates,
and denote $a_q$ its area relative to the total surface area. For
example, $a_{80\%}$ is the relative area on the substrate from
which, on average, 80\% of the signal originates. If the HS is
strong and highly localized, then $a_q$ should be small, even for
$q$ close to 1 (most of the signal originates from a very small
region). Note that by definition, $a_{100\%}=100\%$. General
expressions for $a_q$ are given in Appendix \ref{AppA} as a function
of the HS parameters. Examples of the relative areas of $q$-HS for
our example are given in Table \ref{Tabaq}. These numbers confirm
the figure quoted in the previous discussion.

\begin{table}
    \caption{\label{Tabaq} Characteristics of $q$-HS's for the example considered here.
    The $q$-HS's are disk-like regions around the HS from which a proportion $q$ of the
    total SERS signal originates. They are defined as $\theta < \theta_q$ (see Fig. \ref{FigLambdaDep}(a) for the definition of $\theta$). $F_q$ is the
    enhancement at the edge ($\theta=\theta_q$) and $a_q$ is the relative area of the $q$-HS.
    See Appendix \ref{AppA} for derivation of these values.}
    \begin{ruledtabular}
        \begin{tabular}{lddd}
          \multicolumn{1}{c}{$q$} & \multicolumn{1}{c}{$F_{\mathrm{max}}/F_q$}
           & \multicolumn{1}{c}{$a_q$} & \multicolumn{1}{c}{$\theta_q$ ($^\circ$)} \\
          \hline
            50\%    & 2.23  & 0.26\%    & 5.8 \\
            80\%    & 6.43  & 0.64\%    & 9.2 \\
            90\%    & 14.3  & 0.97\%    & 11.3 \\
            95\%    & 32.0  & 1.34\%    & 13.3 \\
            98\%    & 92.3  & 1.89\%    & 15.8 \\
            99\%    & 206.0 & 2.37\%    & 17.7 \\
          \end{tabular}
    \end{ruledtabular}
\end{table}

\section{Hot-spot to hot-spot variation}
\label{SecHStoHS}

We have so far focused on describing one single HS. This approach is
relevant to many experimental situations. In particular, for SM-SERS
studies, the experimental conditions are usually adjusted to achieve
exactly this, a maximum of one HS at a time in the scattering
volume. There are however other situations where the signals will
originate from a substrate containing many hot-spots, for example in
a large colloidal aggregate. In such conditions, the single HS model
is still useful since the total signal can in a first approximation
be taken as the sum of signals from independent hot spots. It is
however necessary to understand how the enhancement distributions
may change from one HS to another. This study is also necessary to
understand series of measurements where one single HS may be
measured at any given time, but not necessarily the same one each
time.

\subsection{Incident polarization effects}
\label{SecHSPol}

The first obvious source of variation from one HS to another is
the variation in the orientation of the incident field
polarization. It is well known that gap HS's are highly uniaxial,
\cite{2006PCCPPol} and optimal enhancements are obtained when the
incident field polarization matches the HS axis. In the previous
study of our model HS, we assumed that the incident field was
along the dimer axis, i.e. optimum coupling. In a more general
case, the results depend on both the incident wave-vector
direction and polarization. In a first approximation, the dimer
has a response similar to a dipole \cite{2006PCCPAbs,2006PCCPPol},
and only the angle $\alpha$ between polarization and dimer axis is
important, introducing an additional factor $\cos^4 \alpha$ to the
enhancement. Such an example is shown in Fig. \ref{FigDistr}(a)
for $\alpha=45^\circ$. The corresponding curve $F(\theta)$ is
nearly identical to that obtained for $\alpha=0$, only scaled by a
factor $\cos^4 \alpha=1/4$.

Denoting $F_0(\theta)$ the EF for $\alpha=0$, we then have
$F(\theta,\alpha)=F_0(\theta) \cos^4 \alpha$. We assume $F_0$
follows the probability distribution described previously, a TPD
with parameters $k$, $F_{\mathrm{max}}$, and $A$. In an experiment
where one single HS with a fixed $\alpha$ is observed, then $F$
follows like $F_0$ a TPD with the same $k$, but with
$F_{\mathrm{max}}$ reduced to $F_{\mathrm{max}} \cos^4 \alpha$,
and $A$ to $A\cos^{4k} \alpha$. In many situations, the angle
$\alpha$ is however not fixed. For a random orientation of the
dimer axis, $\alpha$ is also a random variable with probability
$p(\alpha)=(1/2)\sin \alpha$. We can then derive the probability
distribution of $F$ and we find that $F$ again follows a TPD, with
the same parameters $k$ and $F_{\mathrm{max}}$, but where $A$ is
replaced by $A/(1+4k)$.

Overall, the previous arguments show that the incident polarization
effects do not significantly change the conclusions drawn earlier.
The EF distribution still follows a TPD with the same $k$ parameter.
Only small changes to the other two parameters apply, transforming
them into ``effective'' parameters averaged over the incident
polarization directions.

\subsection{Different types of HS}

We now briefly study how the distribution may vary for different
types of gap HS's. Within our model gap HS formed by a dimer, one
could expect the distribution to change with a number of parameters:
 wavelength $\lambda$, separation (gap) between particles $g$, radius of
 particles $a$, optical properties of the metal.
 We carried out calculations of the EF distributions for various combinations
 of these parameters. The first observation is that for all cases
 studied here, the distribution of EF is
 always ``long-tail'' and extremely well-represented by a TPD (in the high EF region),
 as in the example studied previously. Like in many other cases in SERS,
 one could argue that the example of a dimer captures the essence of
 most high enhancement situations, irrespective of the details. The parameters of the TPD for a number of
 representative cases are summarized in Table \ref{TabAllHS}.

\begin{table*}
    \caption{\label{TabAllHS} Example of various gap HS's formed by two closely-spaced metallic
    particles excited with a field polarized along the dimer axis.
    The HS's characteristics depend on a number of parameters: metal (silver or gold), radius $a$ of each particle,
    gap $g$ between particles, and excitation wavelength $\lambda$. The (R) after $\lambda$ indicates that
    this particular wavelength corresponds to the plasmon resonance of the dimer
    (where the maximum enhancements are obtained).
The enhancement distribution can be fitted to a truncated Pareto
distribution with parameters $k$, $F_{\mathrm{max}}$, and $A$. The
sharpness and localization can be further described by the values
of $D$ and $a_q$ (defined in the text and in Appendix A).}
    \begin{ruledtabular}
        \begin{tabular}{clll|clccc}
            \multicolumn{4}{c|}{HS characteristics} & \multicolumn{5}{c}{TPD
            parameters}\\
            Metal & $a$ (nm) & $g$ (nm) & $\lambda$ (nm) $\qquad$ & $k$ & \multicolumn{1}{c}{$F_{\mathrm{max}}$} & $A$ & $D$ & $a_{80\%}$ \\

       \hline \multicolumn{4}{c|}{} & \multicolumn{5}{c}{}\\
            Ag & 25 & 2 & 447 (R)   & 0.135 & $1.9\times 10^{10}$ & 0.075 & 285 & 0.64\% \\
            Ag & 25 & 2 & 467       & 0.152 & $2.3\times 10^{9}$ & 0.086 & 258 & 0.72\% \\
            Ag & 25 & 3 & 429 (R)   & 0.148 & $2.5\times 10^{9}$ & 0.114 & 183 & 1.01\% \\
            Ag & 25 & 3 & 447       & 0.156 & $4.3\times 10^{8}$ & 0.113 & 166 & 1.13\% \\
            Ag & 25 & 1 & 486 (R)   & 0.128 & $6.5\times 10^{11}$ & 0.055 & 512 & 0.35\% \\
            Ag & 12.5 & 1 & 429 (R) & 0.120 & $4.2\times 10^{11}$ & 0.066 & 327 & 0.55\% \\
            Au & 25 & 2 & 558 (R)   & 0.163 & $1.0\times 10^{8}$ & 0.068 & 248 & 0.76\% \\
        \end{tabular}
    \end{ruledtabular}
\end{table*}

The most interesting aspects of this table are briefly discussed:
\begin{itemize}
 \item
 The main difference among HS's appears in the value of $F_{\mathrm{max}}$, which
varies by 3-4 orders of magnitude here. However, this variation only
reflects a global scaling factor on the whole distribution of
enhancement, not on its shape. For example, the average enhancement
scales by approximately the same amount.
 \item
 $F_{\mathrm{max}}$ describes the strength of the HS and it is
 stronger in the following situations:
(i) for silver compared to gold, (ii) when the excitation wavelength
matches the (red-shifted) plasmon resonance of the dimer, (iii) for
smaller gaps, and (iv) for larger curvature around the HS (i.e when
$a$ is smaller). These four aspects manifest themselves here in the
specific context we are studying, but are in fact well-known in the
general picture of field enhancements in metallic nano-structures.
 \item
 The most remarkable feature, however, is in the actual distribution of
 enhancements. The parameters $k$ and $A$ do not vary much
 from one HS to another, even for
relatively large variation of the gap $g$, or changing from silver
to gold (which leads to a change from 0.26 to 2.0 in the imaginary
part of the dielectric function $\epsilon(\lambda)$ at the
wavelengths considered here).
\end{itemize}

To summarize this section, the shape of the enhancement distribution
remains in a first approximation independent of the HS
characteristics. The HS strength is therefore the dominant parameter
in the variability from one HS to another. Hence, the considerations
of the previous sections apply to a wide variety of gap HS's. The
distribution is always extremely skewed (``long-tail'') with a $k$
parameter in the range $\sim 0.12-0.18$, and is simply displaced
towards the lower or higher enhancements region depending on the
actual strength of the HS. This conclusion, we believe, will hold
for most types of gap HS's.

Note finally that one could consider other types of HS's, such as
those formed in a groove, or at sharp corners. These HS's are not as
strong and therefore less likely to be relevant to SM-SERS studies.
The ``long-tail'' nature of the distribution will remain, but the
parameters could change compared to gap HS's. These would require a
separate study.

\subsection{Collection of hot-spots}

As mentioned before, there are experimental situations where a
large number of HS's are present. We will not dwell on this aspect
here since it is less relevant to SM-SERS, but simply make a few
general remarks.

In most situations involving a large number of HS's, the substrate
may be considered as a collection of spatially separated independent
single HS's. The distribution of EF's can then in a first
approximation be obtained by summing the contribution from these
HS's (i.e. summing their respective pdf's). Note however that the
HS's can come from regions which will have in general different
relative areas, and the sum should therefore be weighed to take this
into account. Such a weight factor can modify the $A$ parameters of
the various $p(F)$. The other main difference between the various
$p(F)$ can come from different $F_{\mathrm{max}}$, which may vary,
say by $\sim 3-4$ orders of magnitude depending on the HS strength,
and incident polarization. In order to be quantitative in such a
situation, one would need to assume a distribution of
$F_{\mathrm{max}}$'s and then estimate the sum of the various
$p(F)$'s. If the $F_{\mathrm{max}}$'s are relatively uniform across
HS's, the total EF distribution will simply resemble that of a
single HS with a $k$ parameter in the range $\sim 0.12-0.18$, as
before. If the $F_{\mathrm{max}}$'s vary over several orders of
magnitude, then a decreasing number of HS's contribute to the
distribution at higher enhancements. This will result in a slight
``bending'' of the straight line representing $\log_{10} p(L)$ in
Fig. \ref{FigDistr}(c) in the region of large $L$'s. The
distribution remains highly skewed and similar to a Pareto
distribution, but this could increase the value of the $k$ parameter
in the region of interest (high enhancements).

\section{Statistics of SERS signals}
\label{SecStats}

We now focus on the consequences of the distribution of SERS
enhancements on the signal statistics. We have seen that this
distribution is always highly skewed when HSs are present, and can
be modelled as a truncated Pareto distribution in most cases of
interest. We will therefore again use the same example of HS's as in
the previous section to illustrate our arguments, but the following
considerations are fairly general and apply to any system containing
HS's (and therefore a ``long-tail'' distribution of enhancements).
It is particularly appropriate to single molecule SERS experiments,
where single HS's are usually studied. As before, we consider in the
following a model HS, here with $N_{\mathrm{mol}}$ molecules
randomly positioned on the surface. We assume that a series of SERS
measurements is carried out on this system and analyze the
statistics of the SERS signals; a situation that has been studied
experimentally in the context of SM-SERS in countless opportunities.
To the very best of our knowledge, there has never been a serious
attempt to pin down the physical meaning of the statistics of events
in these experiments by means of a model distribution for the
enhancements at HS's. The next subsections, therefore, present some
of the key results of the analysis using the tools developed so far.

\subsection{Ultra-low concentration}

 Let us first consider the extreme case of
$N_{\mathrm{mol}}=1$. Note that in most SM-SERS experiments, it is
claimed that there is indeed less than one molecule at a time in the
scattering volume. \cite{1997Nie,1997Kneipp} In this case, the SERS
signal from this molecule will only be detected if it is located at
a position of sufficiently high enhancement $F>F_t$, i.e. close to
the HS, where $F_t$ depends on the sensitivity of the detection and
the intrinsic scattering properties of the probe. Let us assume, for
example, that only molecules experiencing an EF of $F_t\approx
F_{\mathrm{max}}/100$ or more can be detected (note that this is a
very optimistic estimate). Then one can see that $p(F>F_t)=a_{99\%}
\approx 2.37 \%$, i.e a signal will be observed for only 1 event in
every 42. Moreover, the intensity of these events will fluctuate by
a factor of around 100, depending on exactly where the observed
molecule is located in the HS. Using a more realistic factor (10
instead of 100), only one SM event in every 100 is detectable.
 Such a behavior was observed experimentally by Nie
{\it et al.} \cite{1997Nie}. Only a small proportion of the colloids
were active, i.e. only the ones where the molecule was by chance at
the HS. In the limit of ultra-low concentrations, it is then
extremely difficult to demonstrate that the number of detected
events scales with concentration due to the worsening statistical
significance. This opens the door to alternative explanations based
on ``rare events'' which are not necessarily SM.

\subsection{Medium concentrations}

\begin{figure}
\centering{
\includegraphics[width=8cm]{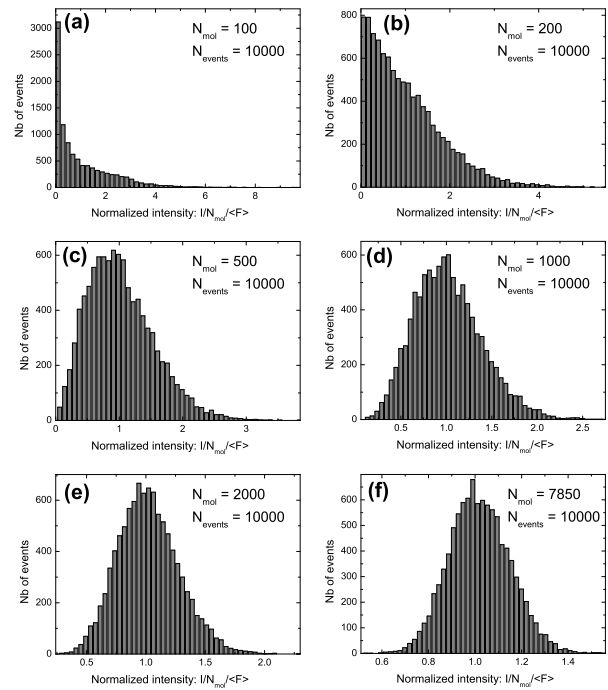}
\caption{Statistics of SERS signals for events consisting of
$N_{\mathrm{mol}}$ randomly adsorbed molecules on a SERS substrate
consisting of the model hot-spot described in Sec. \ref{SecEF}. Each
histogram is calculated from 10000 events. They represent the
distribution of normalized SERS intensity
$I_N=(\sum_{i=1}^{N_{\mathrm{mol}}} F_i)/(N_{\mathrm{mol}}\langle
F\rangle)$. The average value for all is therefore $\langle
I\rangle=1$. The scale for each histogram is adjusted from
$I_{\mathrm{min}}$ to $I_{\mathrm{max}}$ (the bars may not be
visible if there are only a few events). Long-tail distributions are
obtained up to $N_{\mathrm{mol}}=500$ and remain asymmetric up to
$N_{\mathrm{mol}}=2000$.}
 \label{FigHistos}
}
\end{figure}

It is now interesting to study what happens for more than one
molecule on the surface and see whether the fluctuations can be
averaged out by having several molecules instead of one. A simple
approach is to consider $a_{50\%}$, which is the relative area from
which 50\% of the SERS signal originates. If we have on average only
one molecule in this small area (corresponding to
$N_{\mathrm{mol}}\approx 1/a_{50\%}$), then we should typically
expect fluctuations of the order of a factor of $\sim 2$ (depending
whether zero, one, or two molecules are in this area). In our
example, $a_{50\%}\approx 0.26\%$, so $N_{\mathrm{mol}}\approx 250$.
We therefore expect substantial fluctuations even for 250 molecules
on the surface. This is simply due to the nature of the enhancement
distribution.

To understand further the statistics of SERS signals, we would
ideally like to determine the probability distribution of the
total intensity of $N_{\mathrm{mol}}$ randomly distributed
molecules. This is a standard problem in probability theory:
finding the pdf of the sum of $N$ random variables with a known
pdf. Unfortunately, there is no simple analytical expression (to
our knowledge) for a sum of random variables with a truncated
Pareto distribution. The easiest approach then is to compute
numerically a histogram of SERS intensities for a large number of
events of $N_{\mathrm{mol}}$ randomly distributed molecules. This
represents the expected intensity distribution in an experiment
with an average of $N_{\mathrm{mol}}$ molecules on the surface.
Examples of such intensity distributions are shown in Fig.
\ref{FigHistos} for our model HS and various values of
$N_{\mathrm{mol}}$. The $x$ axis in these histograms shows the
intensity normalized with respect to the number of molecules and
to the expected average signal per molecule $\langle F\rangle$. A
value of 1 therefore corresponds to the average signal. The
smallest and largest value on the $x$ axis corresponds to the
intensities of the smallest and largest event (if there is only
one such event, the bar in the histogram is too small to be
visible).

We can make a number of remarks on these distributions:
\begin{itemize}
\item The distribution remains skewed (``long-tail'') even for large numbers of
adsorbed molecules (at least 500 molecules). Large intensity
fluctuations should therefore be expected in this range. As
discussed before, this is due to the strong localization of the HS.
The observed SERS intensity for 500 molecules has in fact a strong
contribution from the few molecules (possibly one or two) closest to
the HS. It is therefore not surprising to have fluctuations
depending on the exact number and position of these few molecules.
This is probably one of the most widespread experimental situations
in SERS: low (but not ultra-low) concentrations.
\item
 Even for 1000 or 2000 molecules, there is
still an asymmetry in the distribution, i.e a long tail with a
small probability of large events. This should be viewed as an
intrinsic property of any HS-containing SERS substrate. The
Pareto-like distribution of enhancements will always lead to a
distribution of SERS intensities with a high-intensity tail,
similar to a {\it lognormal} or {\it Gamma} distribution. This is
true even for relatively large numbers of molecules and would only
disappear for extremely large values (for example
$N_{\mathrm{mol}}=7850$ in the figure) where the central-limit
theorem prediction of a Gaussian distribution is recovered. \item
For small numbers of molecules, for example,
$N_{\mathrm{mol}}=100$, there are many events with intensities
below the detection limit. They correspond to events where no
molecule is close enough to the HS to experience a sufficient
enhancement. At such concentrations, the SERS signals originate
essentially from a single molecule (the other 99 have a much
smaller contribution). SM-SERS is therefore possible (and in fact
much more practical) at concentrations much larger than typically
used in ultra-low concentration studies. Decreasing further the
concentration, as often done in SM-SERS experiments, will only
decrease the probability of SM-SERS events and make the analysis
more difficult and the statistics unreliable. The only issue at
intermediate concentrations is to be able to prove that the
signals are indeed SM signals. A bi-analyte technique as described
in Ref. \onlinecite{2006JPCBBiASERS}, or an equivalent method,
must then be used to verify the SM nature of the signals.

\end{itemize}

It is clear from these examples that intensity fluctuations are an
intrinsic property of SERS in most substrates with HS's. It is
important to realize that they are not only a consequence of the
variability of the SERS substrates itself. Even in a solution of
exactly identical metallic dimers for example, fluctuations will
arise for $N_{\mathrm{mol}} < 1000$ molecules per colloids simply as
a result of the random adsorption combined with the highly skewed
(``long-tail'') distribution of SERS enhancements. {\it Intensity
fluctuations alone cannot be invoked as a proof of SM-SERS}.

\subsection{High concentrations}

It is interesting to see whether it is possible to minimize or
eliminate these fluctuations. If one could have a large amount of
molecules on the surface, then the observed SERS signal should be
$N_{\mathrm{mol}} \langle F\rangle$ with little fluctuations. An
interesting question is how many molecules we need to remove the
fluctuations. We could try to use the central limit theorem to
answer this question but the distribution is so skewed that it may
not apply except for very large $N$ (slow convergence); a property
that is clearly hinted at in Fig. \ref{FigHistos}. If it did
apply, it would tell us that for sufficiently large $N$, the
fluctuations around $N\langle F\rangle$ are of the order of
$\sigma \sqrt{N}$. If we define the term ``small fluctuations'' by
the condition $\sigma/(\sqrt{N} \langle F\rangle) \le 10\%$, this
then requires to take $N \ge 100 (\sigma/\langle F\rangle)^2$. An
important aspect of long-tail distributions is that the standard
deviation, $\sigma$, is very large, even much larger than $\langle
F\rangle$ in many cases (see Appendix \ref{AppA}). In our example,
we therefore need $N>13000$ !! This very large number again
demonstrates the highly unusual nature of distribution we are
dealing with for HS's.

Now it is worth remembering that in SERS, we cannot in general have
as many molecules as we want on a substrate's surface. If we assume
that only the first layer is active\cite{Otto}, then the maximum
number of molecules depends on the surface area of one molecule
compared to that of the whole substrate. If we consider molecules
with surface coverage of $\sim 1$\,nm$^2$ (reasonable for dyes), and
our model HS, which has a total surface of $S=7850$\,nm$^2$, then
$N_{\mathrm{max}}$ is of the order of 8000. The fluctuations are
then reasonably small but still present (see Fig.
\ref{FigHistos}(f)). Moreover, for many dyes in colloidal solutions,
the maximum concentration can be limited by induced aggregation of
colloids and fluctuations may then be unavoidable, even for the
largest allowed concentrations. The SERS fluctuations can then be
viewed as an unavoidable (and natural) consequence of the long-tail
distribution of enhancements around HS's.

\subsection{How to interpret ``Poisson distributions of
SERS intensities''?}

\begin{figure}
\centering{
\includegraphics[width=8cm]{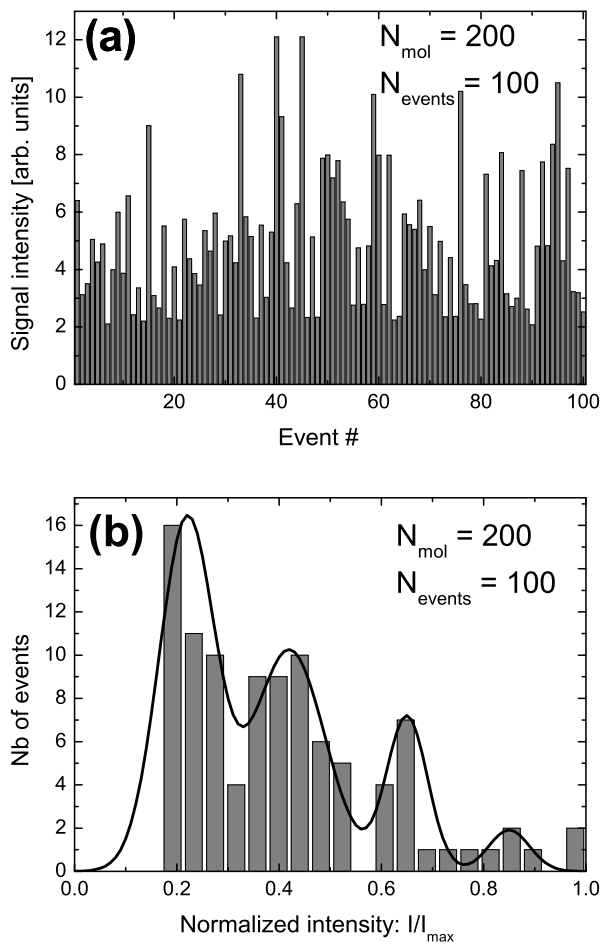}
}
 \caption{Illustration of how a Poisson-like distribution is
 obtained from a random sample of 100 events for 200 randomly
 adsorbed molecules. (a) Fluctuations of the SERS intensity, where a
 background intensity of 2 has been added to resemble a typical
 SM-SERS experiment. (b) Histogram of intensities of the 100 events,
 together with a ``Poisson'' fit with Gaussian broadening. The
 ``Poisson'' parameter in this fit is 0.8. The Poisson-like nature
 of this histogram is artificial and due to the poor sampling (100
 events). Taking 10000 events results in the histogram of Fig.
 \ref{FigHistos}(b).
 }
 \label{FigPoisson}
\end{figure}

A Poisson distribution of SERS intensities is often put forward as
an argument for SERS SM-signals.
\cite{1997Kneipp,2006APLSMSERS,2006PRBSMTERS} It was indeed one of
the main argument of one of the first reports on SM-SERS.
\cite{1997Kneipp} It is in general argued that the peaks in the
Poisson distribution are a result of seeing none (background
signal), one, two, or three molecules during a given SERS event. The
average number of molecules can then be inferred from the relative
intensities of these Poisson peaks. The conditions for this argument
to be true have not been emphasized enough in our opinion:
\begin{itemize}
\item The argument is only valid if each single molecule gives
exactly the same SERS signal (within less than a factor of 2 to
differentiate between one strong molecule and two weak molecules).
This is highly unlikely in SERS for many reasons: non-uniform
exciting beam, changing orientation of substrate and/or molecule
with respect to incident polarization, and most importantly the
large spread in the distribution of SERS enhancements.
 \item
  One
could argue that there is a saturation mechanism, for example
photo-bleaching of the molecule, which guarantees that every
single molecule emits more or less the same signal before being
destroyed. However, the photo-bleaching mechanisms under SERS
conditions are still not well understood and it is not clear
whether it would result in such a perfect saturation of the SERS
signal from a single molecule (which follows a different
enhancement factor than absorption or fluorescence).
 \item
  Even in
the presence of a suitable saturation mechanism, the concentration
inferred from the Poisson distribution would be much smaller than
the experimental molecular concentration. This is because of the
strong spatial localization of the HS: the chances of a molecule
being at a HS are small, and the molecule is undetetectable if it
is outside the HS.
 \item
  If the concentration inferred from the
Poisson distribution matches the experimental concentration, as in
Ref. \onlinecite{1997Kneipp}, then it means that every single
molecule is adsorbed at the HS. This is very unlikely in general. It
has been suggested that laser forces attracting the molecule towards
the HS's\cite{2006KallFD,2002PRLXu} could provide an additional
mechanism to improve the chances of seeing molecules at HS's, but
this remains very speculative at the moment. In most studied
situations it will be the surface chemistry and molecular
interaction of the molecule with the surface that determines the way
the analyte is spread over the surface.
\end{itemize}

We here propose a simpler, and arguably more credible,
interpretation of these Poisson distributions. Let us first note
that they are always obtained {\it from a small number of events},
100 in general. \cite{1997Kneipp,2006APLSMSERS,2006PRBSMTERS} It is
in fact a property of long-tail distributions that most random
samples of $\sim 100$ events exhibit peaks similar to a Poisson
distribution. {\it This is true for any long-tail distributions},
such as lognormal distributions or those shown for example in Fig.
\ref{FigHistos}(a-c) for $N_{\mathrm{mol}}\le 500$. To illustrate
this, we show in Fig. \ref{FigPoisson} an example of the SERS
intensities of 100 events for 200 molecules randomly distributed on
the surface, along with a ``Poisson'' fit to the data. This fit is
misleading, because it is only a result of the poor sample size used
for the histograms (100 events). In fact, when using 10000 events,
we simply obtain the histogram of Fig. \ref{FigHistos}(b), which is
a ``standard'' long-tail distribution. This type of ``Poisson'' fits
is therefore not a proof of SM-SERS, but rather illustrates the wide
range of enhancements in a typical SERS experiments with HS's and is
another manifestation of the peculiar nature of the distribution
present in this problem.

\subsection{Other experimental consequences of the EF distribution}

We briefly discuss here two additional effects related to the
enhancement distribution: vibrational
pumping\cite{Kneipp1,Brolo,Haslett,RobFaraday,RobJPCB,RobJPCB2,RobJPCB3}
and photo-bleaching under SERS conditions.

We have recently proposed a clear demonstration of the existence of
vibrational pumping under SERS conditions.\cite{RobJPCB2,RobJPCB3}
This effect can in principle be used to extract the SERS
cross-section of the analyte, and therefore the SERS enhancement.
For a non-uniform distribution of enhancements, we have shown
\cite{RobJPCB2,RobJPCB3} that what is measured is not the average
enhancement $\langle F\rangle$, but the pumping enhancement given by
$F_p=\langle F^2\rangle/\langle F\rangle$ and argued that it was a
good (under)estimate of the maximum enhancement $F_{\mathrm{max}}$.
Using the truncated Pareto distribution of enhancement, we can give
a more accurate estimate. Using Eq. (\ref{EqnF2}) and the fact that
$k$ is small, we deduce that the pumping enhancement, $F_p$, is
approximately a factor of $\sim 2$ smaller than the maximum
enhancement.

Another phenomenon where the distribution of enhancement is likely
to play an important role is photo-bleaching under SERS conditions.
For conventional photo-bleaching, the number of molecules (and
therefore the intensity) decays as $n(t)=n_0 \exp(-\Gamma t)$ where
the decay rate $\Gamma$ depends on the laser intensity, in a first
approximation as $\Gamma=\beta I_L$. The mechanisms of
photo-bleaching under SERS conditions are still not well resolved,
but it is reasonable to assume that $\Gamma$ will also depend on the
enhancement factor $F$ experienced by the molecule. Two simple
assumptions are $\Gamma(F)=\beta F I_L$ (proportional to SERS EF),
or $\Gamma(F)=\beta \sqrt{F} I_L$ (proportional to absorption or
local field intensity enhancement). This dependence would result in
both cases in a non-exponential decay of the SERS intensity, which
can be estimated from:
\begin{equation}
I(t)=N_{\mathrm{mol}} \int  p(F) F e^{-\Gamma(F) t} dF.
\end{equation}
This equation in fact provides a starting point for using
photo-bleaching as a way of measuring EF distributions
experimentally. Combined with theoretical estimations of EF
distributions, it also provides a chance of studying the
photo-bleaching mechanism itself, i.e. determining the dependence
$\Gamma(F)$ and its physical origin. As an example, for a TPD as
considered here, the expected decay of $I(t)$ (for long times) are
approximately power laws scaling as $1/t^{1-k}$ if $\Gamma(F)=\beta
F I_L$ or as $1/t^{2-2k}$ if $\Gamma(F)=\beta \sqrt{F} I_L$.

\section{Conclusion}
The main purpose of this paper was to study the distribution of
enhancements for HS's in SERS and to understand some very general
properties. We have shown that most problems on the statistics of
signals coming from HS's and SM-SERS in general can be tracked down
to the existence of a {\it very skewed} (long-tail) distribution of
enhancements. Within this framework a series of natural consequences
and explanations arise, among them: $(i)$ the role and meaning of
fluctuations at ultra-low concentrations, $(ii)$ a proper
justification of SM signals at medium concentrations (which is a
justification for the bi-analyte method for SM-SERS developed in
Ref. \onlinecite{2006JPCBBiASERS} at the same time), and $(iii)$ the
observation of probability distribution oscillations
(``Poisson''-like), which are not by themselves a proof of SM-SERS
but have their origin in the very peculiar characteristics of
long-tail distributions. We also explored very briefly some of the
consequences that the distribution of enhancement would have in
vibrational pumping and photobleaching under SERS conditions. This
latter predictions will have to await for experimental confirmation
before further development is justified.

Overall, the approach taken here has the advantage of its
``universal'' nature in view of the fact that a power law
distribution of enhancements is most likely to exist in SM-SERS
substrates with HS's, independent of the exact details of the
situation. This provides, therefore, not only a general
understanding and a theoretical framework (in a field that has been
plagued with diverging interpretations), but also a powerful
phenomenological tool to describe the statistics of SM-SERS signals
with a minimum set of parameters.

\appendix
\section{}
\label{AppA}

We here consider the model distribution of enhancement for a single
HS as defined by a truncated Pareto distribution with parameters
$k$, $A$, and $F_{\mathrm{max}}$. The parameters obtained in the
example of Sec. \ref{SecEF} will be used for illustration:
$k=0.135$, $A=0.075$, and $F_{\mathrm{max}}=1.9 \times 10^{10}$.

The pdf of the SERS enhancement factors, $F$, is then given by:
\begin{equation}
p(F)=AF^{-(1+k)}\quad\mathrm{for}\quad F_{\mathrm{min}} < F <
F_{\mathrm{max}}.
\end{equation}

We have introduced here a minimum value for $F$, $F_{\mathrm{min}}$,
which is required for $p(F)$ to be a valid pdf. $F_{\mathrm{min}}$
can be deduced from the parameters using the normalization condition
for the pdf,
\begin{equation}
\int_{F_{\mathrm{min}}}^{F_{\mathrm{max}}} p(F) dF=1,
\end{equation}
which implies that
\begin{equation}
F_{\mathrm{min}}=\exp\left(-\frac{1}{k}\ln\left(\frac{k}{A}\right)\right).
\label{EqnFmin}
\end{equation}
Note that $F_{\mathrm{min}}$ is only introduced for mathematical
consistency in the definition of $p(F)$, it has no physical meaning
and will be irrelevant in all predictions. Its value is typically
very small (and in fact much smaller than the real physical minimum
enhancement on the surface). For our example,
$F_{\mathrm{min}}\approx 7.6\times 10^{-3}$.

Equivalent representations of the EF distribution can be given in
terms of $L=\log_{10} F$, for which
\begin{equation}
p(L)=A\ln (10) 10^{-kL},
\end{equation}
or in terms of the local field intensity enhancement $M=\sqrt{F}$,
for which
\begin{equation}
p(M)=2A M^{-(1+2k)}.
\end{equation}

Other quantities of interest that can be derived are, firstly the
average enhancement, $\langle F\rangle$, given by
\begin{equation}
\langle F\rangle=\int_{F_{\mathrm{min}}}^{F_{\mathrm{max}}} F p(F)
dF\approx \frac{A F_{\mathrm{max}}^{1-k}}{1-k} \label{EqnAveF1}.
\end{equation}
It can also be written as:
\begin{equation}
\langle F\rangle=\frac{F_{\mathrm{max}}}{D}
\quad\mathrm{with}\quad D=\frac{1-k}{A}F_{\mathrm{max}}^k
\label{EqnAveF2}
\end{equation}
$D$ characterizes the strength of the HS with respect to the average
enhancements, and is usually quite large ($D=285$ in our example).
This is a defining characteristics of all long-tail probability
distributions.

One can also show that:
\begin{equation}
\langle F^2\rangle=c F_{\mathrm{max}}\langle
F\rangle\quad\mathrm{with}\quad c=\frac{1-k}{2-k} \label{EqnF2}
\end{equation}
Note that $c \approx 1/2$ when $k$ is small, which is often the
case ($c\approx 1/2.16$ in our example). Moreover, using Eq.
(\ref{EqnAveF2}) and the fact that $D\gg 1$, we derive the
standard deviation $\sigma$ of $F$:
\begin{equation}
\sigma \approx \sqrt{cD} \langle F\rangle.
\end{equation}
$\sigma$ is therefore much larger than the average $\langle
F\rangle$ (by a factor $\sqrt{cD}\approx 11.4$ in our example),
which is also an important characteristic of long-tail
distributions.

We now focus on the properties of a $q$-HS defined as the region
from which a proportion $q$ of the total SERS signal originates. To
determine its relative area $a_q$, we first find $F_q$, smallest
enhancement in the $q$-HS (at its edge). It can be obtained from the
condition:
 \begin{equation}
\int_{F_q}^{F_{\mathrm{max}}} F p(F) dF=q\langle F\rangle,
 \end{equation}
which leads to
\begin{equation}
F_q=\left(1-q\right)^{\frac{1}{1-k}} F_{\mathrm{max}}.
\end{equation}
Because the center of the $q$-HS corresponds to the place of maximum
enhancement, and because the enhancements decay monotonically away
from the center of the HS, one can see that $a_q$ is then simply
obtained from:
\begin{equation}
a_q=\int_{F_q}^{F_{\mathrm{max}}} p(F)
dF=\frac{AF_{\mathrm{max}}^{-k}}{k}\left[
\left(1-q\right)^{-\frac{k}{1-k}} -1 \right]
\end{equation}
For a HS on a sphere (part of a dimer), we can also deduce the
half-angle $\theta_q$ defining a $q$-HS:
\begin{equation}
\theta_q = \cos^{-1} (2a_q-1)
\end{equation}

\newpage
\printtables

\newpage
\printfigures


\begin{thebibliography}{}
\bibitem{Pareto2} The Pareto distribution is associated with the ``Pareto principle'', which states that
20\% of a population owns 80\% of the wealth.
\end{thebibliography}
\end{document}